\documentclass[aps,prl,twocolumn,nofootinbib,superscriptaddress,showpacs,showkeys]{revtex4-1}

\usepackage{graphicx}
\usepackage{amsmath, amsxtra, amssymb, mathtools, isomath, nccmath, xspace, siunitx}
\usepackage{color}
\usepackage[utf8]{inputenc}
\usepackage[british]{babel}
\usepackage{hyperref}
\usepackage{sidecap}
\usepackage{natbib}
\usepackage{dsfont}
\usepackage{xcolor}

\newcommand{\ket}[1]{\ensuremath{\lvert #1 \rangle}\xspace}%
\newcommand{\bra}[1]{\ensuremath{\langle #1 \rvert}\xspace}%
\newcommand{\Rb}{\ensuremath{^{87}\text{Rb}}\xspace}%
\newcommand{\alat}{\ensuremath{a_{\text{lat}}}\xspace}%
\newcommand{\gtwo}{\ensuremath{g^{(2)}}\xspace}%

\newcommand{\unaryminus}{\scalebox{0.75}[1.0]{\( - \)}}

\renewcommand{\theequation}{S\arabic{equation}}

\sidecaptionvpos{figure}{h}

\long\def\symbolfootnote[#1]#2{\begingroup%
\def\thefootnote{\fnsymbol{footnote}}\footnotetext[#1]{#2}\endgroup}

\begin{document}
\title{\bf{Quantum gas microscopy of Rydberg macrodimers}}


\author{Simon Hollerith}
\affiliation{Max-Planck-Institut f\"{u}r Quantenoptik, 85748 Garching, Germany}

\author{Johannes Zeiher}
\thanks{present address: Department of Physics, University of California, Berkeley, California 94720, USA}

\author{Jun Rui}
\email[]{Jun.Rui@mpq.mpg.de}
\affiliation{Max-Planck-Institut f\"{u}r Quantenoptik, 85748 Garching, Germany}

\author{Antonio Rubio-Abadal}
\affiliation{Max-Planck-Institut f\"{u}r Quantenoptik, 85748 Garching, Germany}

\author{Valentin Walther}
\affiliation{Department of Physics and Astronomy, Aarhus University, DK 8000 Aarhus C, Denmark}
\author{Thomas Pohl}
\affiliation{Department of Physics and Astronomy, Aarhus University, DK 8000 Aarhus C, Denmark}
%
\author{Dan M. Stamper-Kurn}
\affiliation{Department of Physics, University of California, Berkeley, CA 94720, USA}

\author{Immanuel Bloch}%
\affiliation{Max-Planck-Institut f\"{u}r Quantenoptik, 85748 Garching, Germany}
\affiliation{Fakult\"{a}t f\"{u}r Physik, Ludwig-Maximilians-Universit\"{a}t M\"{u}nchen, 80799 M\"{u}nchen, Germany}%
\author{Christian Gross}%
\affiliation{Max-Planck-Institut f\"{u}r Quantenoptik, 85748 Garching, Germany}


\date{\today}


\begin{abstract}

The sub-nanoscale size of typical diatomic molecules hinders direct optical access to their constituents.
Rydberg macrodimers --- bound states of two highly-excited Rydberg atoms --- feature interatomic distances easily exceeding optical wavelengths.
Here we report the direct microscopic observation and detailed characterization of such molecules in a gas of ultracold atoms in an optical lattice.
The bond length of about $\SI{0.7}{\micro\meter}$, comparable to the size of small bacteria, matches the diagonal distance of the lattice.
By exciting pairs in the initial two-dimensional atom array, we resolve more than 50 vibrational resonances.
Using our spatially resolved detection, we observe the macrodimers by correlated atom loss
and demonstrate control of the molecular alignment by the choice of the vibrational state.
Our results allow for rigorous testing of Rydberg interaction potentials and highlight the potential of quantum gas microscopy for molecular physics.

\end{abstract}
\maketitle

\begin{figure}[htb]
  \centering
  \includegraphics[width=0.5\textwidth]{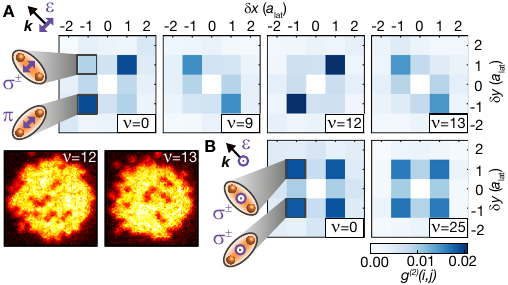}%
  \caption{\label{fig:3}
  \textbf{Macrodimer imaging and molecular orientation on the lattice.}
  (\textbf{A}) Evaluating $\gtwo(i,j)$ after illuminating the atoms with UV light polarized in the atomic plane for various macrodimer lines $\nu$, reveals a significant directionality in the excitation rate.
  Whereas for even vibrational states, the excitation rate is stronger along the polarization of the excitation laser, odd states favor molecule formation perpendicular to the polarization.
  For both cases, we show an exemplary image from our microscope.
  The origin of the alternating molecular orientation is the polarization of the light with respect to the quantization axis of the dimer.
  The light is $\pi$-polarized with both axes aligned and $\sigma^\pm$-polarized in the perpendicular case.
  (\textbf{B}) For out-of-plane polarization, there is no longer a preferred direction and we observe equal correlations for both diagonal directions.
}
\end{figure}

A quantitative determination of the structure of molecules is an essential goal of physical chemistry and is crucial to reveal and understand their properties.
The high level of quantum control and the ultracold temperatures achieved in atomic physics provide novel tools to study molecules and their structure~\cite{Julliene_Review06}.
Prominent examples include the observations of weakly bound Feshbach molecules~\cite{regal_creation_2003,moses_creation_2015}, the controlled photoassociation of individual molecules in a microtrap~\cite{liu_building_2018}, molecules comprising ground-state atoms bound to a highly excited Rydberg atom~\cite{gallagher_1994,Bendkowsky2009,Shaffer2018} or pure long-range molecules~\cite{pure_long_range_1,pure_long_range_2} which are bound purely electrostatically.
The binding mechanism of the latter, where the electron orbitals of the constituents do not overlap, has also been predicted to exist between two Rydberg atoms.
These ``Rydberg 	macrodimers"~\cite{Boisseau2002,Stanojevic2006,Theory_symmetries,CS_calculations_2006, schwettmann_analysis_2007} are truly remarkable in their macroscopic bond lengths, 10\,000 times larger than usual diatomic molecules and thus reaching typical interparticle distances in magneto-optical traps~\cite{Sassmannshausen2016}, optical lattices~\cite{Bakr18} or tweezers~\cite{Bernien2017,Labuhn2016a}.
Their enormous size not only enables direct optical access to individual constituents, but also allows for the controlled binding of two atoms optically pinned at the correct distance.
First signatures of Rydberg macrodimers have been observed in systems of laser-cooled atoms by spectroscopy~\cite{Sassmannshausen2016} and pulsed-field ionization~\cite{Overstreet2009}, but a vibrationally and spatially resolved detection has been lacking so far.\\
Here, we present a precise study of Rydberg macrodimers starting from ground-state $^{87}\textrm{Rb}$ atoms deterministically arranged in an optical lattice.
We probe the vibrational levels by two-photon spectroscopy, resolving more than 50 excited vibrational resonances.
The observed spectrum agrees quantitatively with \emph{ab initio} calculations of Rydberg interaction potentials~\cite{Weber2017,Sibalic2017,Deiglmayr2016}, providing a stringent test for their accuracy.
Using the site-resolved detection and single-atom sensitivity of our quantum gas microscope~\cite{Sherson2010,bakr_probing_2010}, we identify the macrodimer signal microscopically as a loss of pairs of ground state atoms at a distance of a bond length.
Furthermore, we control the spatial orientation of the photo-associated molecules by the parity of the vibrational wave function and the polarization of the excitation laser.\\
\begin{figure*}
\centering
\includegraphics[width=1.9\columnwidth]{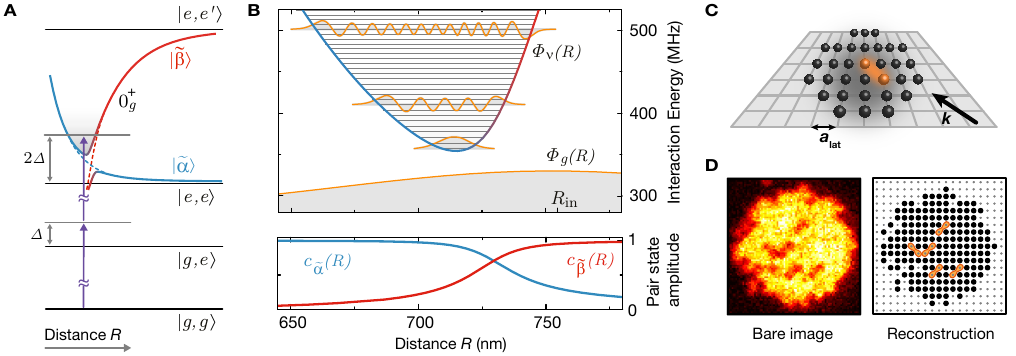}
  \caption{ \label{fig:1}
  \textbf{Schematic of the experiment.}
  (\textbf{A}) The two-photon and two-atom excitation from the ground state $\ket{g,g}$ occurs off-resonantly with detuning $\Delta$ via intermediate states where one atom is excited to the Rydberg state $\ket{e} \equiv |35P_{1/2}\rangle$.
  The avoided crossing of two coupled Rydberg interaction potentials $\ket{\widetilde \alpha}$ and $\ket{\widetilde \beta}$ (blue and red solid lines) leads to the formation of a binding potential, whose molecular states can be laser excited if $2\Delta$ matches the interaction shift from the asymptotic pair state $\ket{e,e}$.
  (\textbf{B}) The potential well hosts bound states (gray horizontal lines) with an energy spacing of around $6\,$MHz.
  The vibrational wave functions, indicated in orange, are much narrower in the internuclear distance $R$ than the rms width of $\sqrt{2}\sigma_{\text{lat}}$ of the initial relative wavepacket $\Phi_g(R)$ in the lattice.
The lower panel shows the amplitudes $c_{\widetilde\alpha,\widetilde\beta}(R)$ resulting from a spatial dependent decomposition of the binding potential.
  (\textbf{C}) The atoms are initially arranged in a regular grid of lattice sites with a lattice constant $\alat=532\,$nm and illuminated by the UV laser with wavevector $\textbf{k}$.
  (\textbf{D}) An exemplary fluorescence image of an atomic cloud illustrates the correlated losses due to molecule formation (left), indicated as orange symbols in the reconstructed image (right).
 }
\end{figure*}
While Rydberg interaction potentials are of van-der-Waals type for asymptotically large interatomic separations $R$~\cite{Saffman2010}, the situation is more complicated at smaller distances.
In a generic situation, repulsive interactions may increase the energy of a lower lying pair state $\ket{\widetilde \alpha}$, while attractive interactions decrease the energy of a higher lying one $\ket{\widetilde \beta}$.
At some specific distance the two potentials become degenerate and any finite coupling between them opens a gap (see Fig.~\ref{fig:1}\,A), which is large when $\ket{\widetilde \alpha}$ and $\ket{\widetilde \beta}$ contain significant amplitudes of mutually dipole-dipole coupled states.
The resulting potential minimum hosts a series of bound macrodimer states $\Phi_{\nu}(R)$~\cite{Boisseau2002, Stanojevic2006, Sassmannshausen2016}, where $\nu$ denotes the vibrational quantum number of the nuclear motion.
In our experiment we choose such an avoided crossing of two $0^{+}_g$ potentials with gerade symmetry and zero angular momentum projection on the interatomic axis~\cite{Julliene_Review06,Weber2017}. 
For large distances, the selected pair states transform into the states $\ket{\widetilde \alpha} \rightarrow |e,e\rangle \equiv |35P_{1/2}, 35P_{1/2}\rangle$ and $\ket{\widetilde \beta} \rightarrow |e,e^\prime\rangle \equiv |35P_{1/2},35P_{3/2}\rangle$~\cite{SI}, which can be optically coupled from the ground state $|g,g\rangle$ by a two photon transition. Here, $|g\rangle = |5S_{1/2} F=2, m_F=0\rangle$ and the bond length of this macrodimer state is predicted to be $720\,$nm, close to the diagonal lattice spacing of $R_{\textrm{in}} = 752\,$nm in our optical lattice.
With the atoms being initially prepared in the motional ground state of the optical lattice, this coincidence of length scales results in a strong optical coupling due to the large wave function overlap.\\
Our experiments started with a two-dimensional atomic Mott insulator of $\Rb$ with a lattice filling of $94(1)\%$ in the atomic ground state.
The atoms were pinned in a deep optical lattice with a rms width $\sigma_{\text{lat}} = 68\,$nm of the motional ground state in the atomic plane (see Fig.~\ref{fig:1}\,B) with a temperature below the on-site trapping frequency.
The molecules were photoassociated by an ultraviolet (UV) Rydberg excitation laser at a wavelength of $298\,$nm propagating along the diagonal direction of the optical lattice, with linear polarization aligned either in or orthogonal to the lattice plane.
Our typical optical Rabi coupling from the state $\ket{g}$ to $\ket{e}$ was $\Omega/2\pi=1.2(1)\,$MHz.
We detected the excited macrodimers as missing pairs of ground state atoms as these are ejected from the optical lattice very efficiently due to the release of kinetic energy in the macrodimer decay.
The remaining ground state atoms were then imaged with near-unity fidelity using a quantum gas microscope, see Fig.~\ref{fig:1}\,D.\\
\begin{figure*}
\centering
\includegraphics[width=1.45\columnwidth]{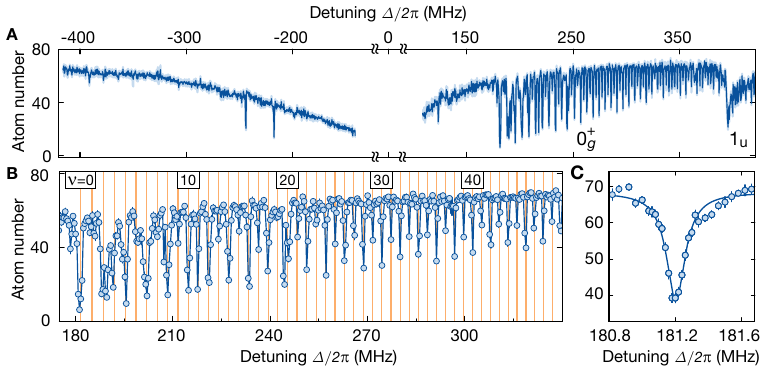}%
  \caption{ \label{fig:2}
  \textbf{High-resolution spectroscopy detuned from the Rydberg resonance.}
  (\textbf{A}) Wide range spectroscopy signal.
  We find no signatures of bound states on the interaction-broadened red-detuned side of the Rydberg resonance (the two isolated features are due to lattice-induced Raman resonances~\cite{SI}).
  On the blue-detuned side, we observe clear dips with regular spacing and alternating line strength due to the coupling to macrodimers.
  For very large detuning a second series of molecular bound states belonging to a $1_u$ potential becomes resonant.
  (\textbf{B}) Zooming into the frequency region between $180$ and $320\,$MHz reveals a spacing of the vibrational resonances of around $3\,$MHz, which slightly decreases for higher vibrational states. 
  We find very good agreement with the theoretical predictions (orange lines), but the line strengths are saturated in the experiment (cf. fig.~S8 in~\cite{SI} for a non-saturated measurement). 
  (\textbf{C}) High-resolution spectroscopy of the lowest vibrational level.
  All error bars on the data points denote one standard error of the mean (s.e.m.).
  }
\end{figure*}
We first aimed at identifying the presence of the bound macrodimers by their spectroscopic fingerprint.
To this end, we illuminated the atomic ensemble for $t_{\textrm{UV}}=100\,$ms with the UV light polarized in the atomic plane for varying detuning $\Delta$ from the bare Rydberg state $\ket{e}$.
In each step, we swept the optical frequency of the excitation laser by $480\,$kHz during the illumination time to ensure coverage of the full spectral region between neighboring data points.
We observe the coupling to the bare Rydberg state $\ket{e}$ as a very broad saturated loss resonance, see Fig.~\ref{fig:2}.
At negative detunings, the resonance features an asymmetric broadening due to coupling to attractively interacting pair-state potentials~\cite{asymmetric_broad_weidemueller}.
At positive detunings around $\Delta/2\pi = 180\,$MHz, the first macrodimer bound state becomes two-photon resonant.
A non-saturated high-resolution spectroscopy of this line (see Fig.~\ref{fig:2}\,C) yields a FWHM of $139(5)\,$kHz, which is of the same order as the measured linewidth of the bare Rydberg resonance~\cite{zeiher_many-body_2016}.
For higher vibrational resonances we observed a reduction of the line strength, which we attribute to a combination of increased intermediate-state detuning and reduced overlap of the spatial wave functions.
In addition, we observed a suppression of the excitation to odd vibrational states $\nu$ due to the approximate odd parity symmetry of the molecular wave functions with respect to the equilibrium distance.
This is consistent with the Franck-Condon principle, which predicts the coupling to be proportional to the overlap integrals of the broad initial and the tightly confined final spatial wave functions $\Phi_{g}(R)$ and $\Phi_{\nu}(R)$, see Fig.~\ref{fig:1}\,B.
However, a closer inspection of the experimental data shows that this simple picture needs to be refined.
Repeating the spectroscopy with orthogonal polarization (so the optical electric field oscillates out of the atomic plane) results in a suppression of the line strength of the even lines.\\
Our microscopic access provides direct \emph{in situ} information about the spatial alignment of the associated molecules and valuable insights into the underlying coupling mechanism~\cite{Samboy2011,Optical_coupling_macrodimers,SI}. 
We compared different molecular lines by illuminating the cloud with UV light, resonant with a given vibrational state $\nu$, until the filling of the lattice decreased to roughly $87(1)\%$.
For a quantitative analysis, we evaluated spatially averaged density-density correlations $\gtwo(i,j)=\langle\langle \hat{n}_{k+i,l+j}\hat{n}_{k,l}\rangle-\langle\hat{n}_{k+i,l+j}\rangle\langle\hat{n}_{k,l}\rangle\rangle_{k,l}$ for the measured spatial atom distributions on a region of interest of $9\times9$ lattice sites, see Fig.~\ref{fig:3}.
Correlations show a clear peak at a distance of a lattice diagonal, revealing the bond length of the molecule.
Moreover, we controlled the orientation of the photo-associated molecules by choosing a combination of vibrational quantum number and polarization of the light field.
For even oscillator states $\nu$, the correlations are stronger along the lattice diagonal parallel to the polarization of the excitation light.
The molecular orientation, however, flips when considering odd oscillator states for which the dimers form predominantly along the direction perpendicular to the polarization.\\
The key to understanding this striking alternation in the orientation of the molecules is the interplay of electronic and motional degrees of freedom. 
The Born-Oppenheimer wave function of the macrodimer can be expressed as $\ket{\Psi^\nu_{\mathrm{Mol}}(R)}= \Phi_{\nu}(R)\,\ket{\chi_{el} (R)}$.
Whereas at large distances, the electronic part $\ket{\chi_{el} (R)}=c_{\widetilde \alpha}(R)\ket{\widetilde \alpha}+c_{\widetilde \beta}(R)\ket{\widetilde \beta}$~\cite{Stanojevic2006,Theory_symmetries} is dominated by the state $\ket{\widetilde \beta}$, for short distances the $\ket{\widetilde \alpha}$ contribution dominates (see Fig.~\ref{fig:1}\,B), and this parametric dependence of the electronic potential has to be taken into account in the optical excitation.
The two-photon Rabi coupling $\widetilde\Omega_{\nu}$ from the ground state to the macrodimer states thus splits into two terms.
Neglecting the weak spatial dependence of the two-photon Rabi couplings $\widetilde\Omega_{\widetilde\alpha}(\widetilde\Omega_{\widetilde\beta})$ to the states $\ket{\widetilde\alpha}(\ket{\widetilde\beta})$, we obtain $\widetilde\Omega_{\nu} \approx  \widetilde\Omega_{\widetilde\alpha} f^\nu_{\widetilde \alpha} + \widetilde\Omega_{\widetilde \beta} f^\nu_{\widetilde \beta}$ with the generalized Franck-Condon integrals $f^\nu_{\widetilde \alpha} = \int\Phi_\nu^{\,*}(R)c^*_{\widetilde\alpha}(R)\Phi_{g}(R){\rm d}R$ and $f^\nu_{\widetilde \beta}$~\cite{Samboy2011, SI}.
Because of the change of the pair state amplitudes around the potential minimum (see Fig.~\ref{fig:1}\,B), $f^\nu_{\widetilde \alpha}$ and $f^\nu_{\widetilde \beta}$ have the same sign for even $\nu$ but an opposite sign for odd $\nu$.
The electronic contributions $\widetilde\Omega_{\widetilde \alpha}(\widetilde\Omega_{\widetilde \beta})$ depend on the alignment of the polarization relative to the molecular axis because the states $\ket{\widetilde \alpha}(\ket{\widetilde \beta})$ obey molecular symmetry contraints.
In the case where the axes are parallel, $\widetilde\Omega_{\widetilde \alpha}$ and $\widetilde\Omega_{\widetilde \beta}$ have the same sign and therefore both terms in $\widetilde\Omega_{\nu}$ add constructively for even vibrational states, leading to a dominating signal in  $\gtwo(1,1)$.
For the perpendicular case, the sign of $\widetilde\Omega_{\widetilde \beta}$ flips and constructive interference occurs for odd vibrational states, resulting in a stronger value for $\gtwo(1,\unaryminus1)$.
Moreover, for the measurement with light polarized out of plane (see Fig.~\ref{fig:3}\,A), the spatial signal exhibits isotropic correlations because in that case the polarization is perpendicular to both lattice diagonals.\\
Although most of the line positions of the measured spectrum shown in Fig.~\ref{fig:2} agree with the theoretical model, we find deviations for low-lying oscillator states. 
A finer scan of that region is shown in Fig.~\ref{fig:4}\,A.
These deviations originate from a third pair state $\ket{\widetilde\gamma}$ asymptotically corresponding to the optically uncoupled state $\ket{32\textrm{D}_{3/2},37\textrm{P}_{1/2}}$ intersecting the binding potential around the potential minimum. 
At the degeneracy point, the weak dipole-quadrupole coupling between the intersecting potentials opens another gap energetically comparable to the vibrational energy. 
As a consequence, a separation of the vibrational motion and interatomic interaction is no longer possible.
For the theory used to describe the coarse vibrational structure shown in Fig.~\ref{fig:2}, we only accounted for the crossing formed by the coupling between $\ket{\widetilde \alpha}$ and $\ket{\widetilde \beta}$. 
We extended our theory by allowing for the vibronic coupling between the vibrational modes and the electronic states $\ket{\widetilde \alpha}$,$\ket{\widetilde \beta}$ and $\ket{\widetilde \gamma}$.
The modified eigenenergies are indicated as orange lines in Fig.~\ref{fig:4}\,B.
Using the refined theory, we indeed can identify almost all observed lines.
To confirm the effect of the intersection, we repeated our spectroscopic measurements for the lowest states in the analogous potential for $n=36$, where such an additional crossing is absent.
In this case, we observed a pure harmonic-oscillator-like spectrum in excellent agreement with the calculations. 
The remarkably high sensitivity of the measured line structure to even weak modifications of the interaction potentials underlines the promise of Rydberg macrodimer spectroscopy for benchmarking Rydberg interaction potentials. 
This also holds for Rydberg interactions in the presence of applied magnetic fields, where accurate calculations are more difficult~\cite{SI}.\\
\begin{figure}[htb]
  \centering
  \includegraphics{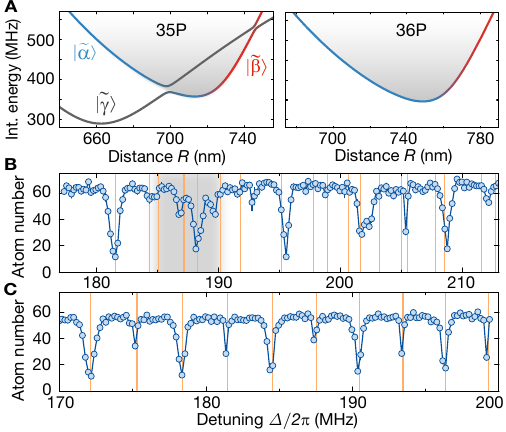}%
  \caption{\label{fig:4}
  \textbf{Breakdown of the Born-Oppenheimer approximation.}
 (\textbf{A}) A closer look at the potential for $n=35$ reveals a $12\,$MHz wide gap, which is absent for $n=36$.
(\textbf{B}) We find a modification of the coarse spectrum discussed in Fig.~\ref{fig:2}\,B in the vicinity of the gap (gray shaded region). A theoretical treatment including the gap (orange lines) allows for an assignment of most of the observed lines.
(\textbf{C}) For $n = 36$, the measured spectrum agrees well with the theoretical expectation. The excitation light was polarized in the atomic plane and all error bars on the data points denote one s.e.m.
  }
\end{figure}
In the future, the coupling to macrodimers could be used to realize quantum gates at well-defined qubit distance or to enhance Rydberg dressing schemes~\cite{Bijnen2015,zeiher_many-body_2016}, where the pair state admixture of the strongly interacting doubly excited state is significantly enhanced compared to the singly-excited intermediate states.
The strongly spatially dependent loss revealed in the correlation measurements and also by modulating the initial atom distribution~\citep{SI} could be used to engineer dissipatively stabilized few- or many-body states~\cite{ates2012,syassen_strong_2008}.
Furthermore, the approach demonstrated here can readily be extended to study multi-atom bound states~\cite{trimers_1,trimers_2}, also in optical tweezers. Finally, bringing the coupling rate to the macrodimers closer to the decay rate of the individual Rydberg atoms may allow for the observation of novel many-body physics arising from spatial constraints and coherent interactions.\\
\begin{acknowledgements}
\textbf{Acknowledgements:}
  We thank all contributors to the open-source programs ``pair interaction'' and ``ARC'' as well as Robin Côté, Bill Phillips, Nikola Šibalić and Johannes Deiglmayr for valuable discussions.
  We acknowledge support by the DNRF through a Niels Bohr Professorship for T.P. and funding by MPG.
  This project has received funding from the European Union’s Horizon 2020 research and innovation programme under grant agreement No. 817482 (PASQuanS) and the European Research Council (ERC) No. 678580 (RyD-QMB) and also from the project No. 319278 (UQUAM) of the Seventh Framework Programme.
  We also acknowledge funding from Deutsche Forschungsgemeinschaft (Project No. BL 574/15-1) within SPP 1929 (GiRyd).\\
  \textbf{Author contributions:} All authors contributed significantly to the work presented in this manuscript.\\
  \textbf{Data availability:} The data that support the plots presented in this paper are publicly available from the Open Access Data Repository of the Max Planck Society (https://edmond.mpdl.mpg.de)~\cite{data} \\
  \textbf{Competing interests:} The authors declare no competing interests.\\

\end{acknowledgements}

\bibliography{RydbergMacrodimers_revtex}

\begin{thebibliography}{41}
\bibitem [{dat()}]{data}%
  \BibitemOpen
  \href@noop {} {}\bibinfo {note} {{The data displayed in the figures of this
  manuscript are available from the Edmond repository of the Max Planck
  Society, doi 10.17617/3.29}}\BibitemShut {NoStop}%
\end{thebibliography}
\setcounter{figure}{0}
\newpage
\section*{Supplementary Materials}
This Supplementary Information contains the calculation of the expected positions of the macrodimer resonances, the wave functions and the optical coupling, additional measurements supporting the statements in the main text and experimental details.
 \section{Calculation of the macrodimer spectrum}
 \begin{figure*}
 \renewcommand\thefigure{S\arabic{figure}}
 \renewcommand\theHfigure{Supplement./thefigure}
  \centering
  \includegraphics{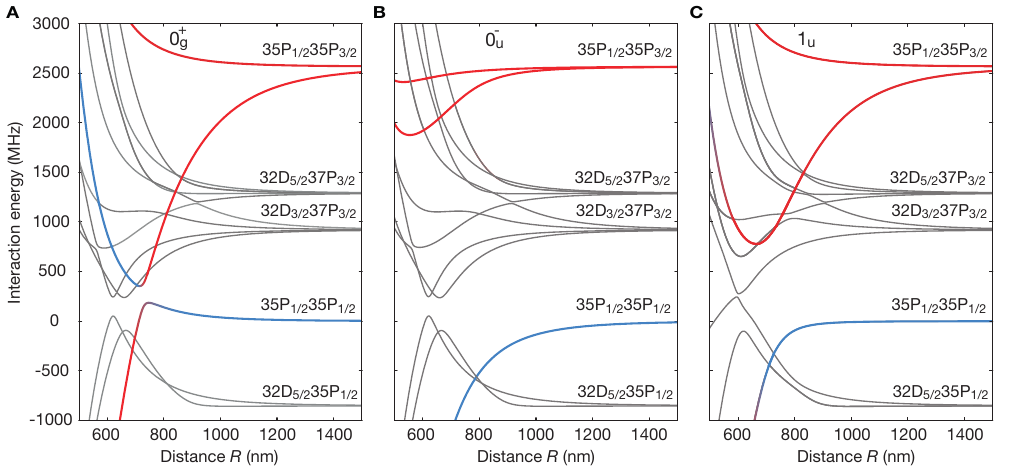}%
  \caption{\label{supfig:5}
  \textbf{Relevant theoretical potential curves.}
  For $\ket{35P_{1/2},35P_{1/2}}$ pair states, there are three relevant interaction branches, which decouple due to the symmetry of the interaction Hamiltonian.
  For the two cases of zero angular momentum projection on the internuclear axis, we find $0^+_g$ potentials with gerade symmetry (\textbf{A}) as well as $0^-_u$ potentials with ungerade symmetry (\textbf{B}).
  If the two angular momenta add up along the molecular axis, there are $1_u$ potentials with ungerade symmetry (\textbf{C}).
  The colored lines highlight the potentials to which atoms are coupled by the light field.
  As in the main text, blue colors denote pair potentials dominated by $\ket{35P_{1/2},35P_{1/2}}$ and red colors denote potentials dominated by $\ket{35P_{1/2},35P_{3/2}}$.
  While the interaction branches at negative interaction energies relative to the asymptotic state $\ket{35P_{1/2},35P_{1/2}}$ are responsible for the broadening of the Rydberg resonance at red detunings, the closed potential wells in the $0^+_g$ and $1_u$ potentials at positive interaction energies host the two series of macrodimer states observed in Fig.~\ref{fig:2}\,A.
  }
\end{figure*}
\begin{figure}[htp]
\renewcommand\thefigure{S\arabic{figure}}
 \renewcommand\theHfigure{Supplement1./thefigure}
  \centering
  \includegraphics{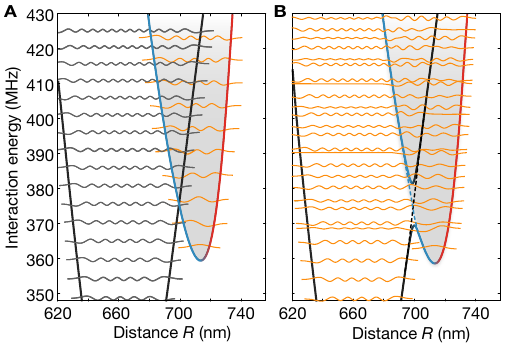}%
  \caption{\label{supfig:8}
  \textbf{Diabatic and vibronic eigenstates.}
  (\textbf{A}) Decoupled vibrational states $ \Phi^{\nu}_{1}(R) $ (orange) and $ \Phi^{\nu}_{2}(R) $ (gray) in the diabatic potentials $V_1(R)$ and $V_2(R)$.
  The discussion of the diabatic states is limited to the eigenstates in the optically coupled potential $V_1(R)$.
  (\textbf{B}) Vibronic eigenstates $|\Psi_{\mathrm{Mol}}^\mu(R) \rangle$ where the nuclear motion cannot be described independently of the electronic degree of freedom anymore.
 Now we plot the adiabatic potentials $\bar{V}_1(R)$ and $\bar{V}_2(R)$ to highlight the presence of the coupling.
  Note that both presented results are different from a calculation of vibrational states in the adiabatic potentials, which does not explain our observations for $n=35$.
  }
\end{figure}

From the discussion of Fig.~\ref{fig:2} and Fig.~\ref{fig:4}\,B, it becomes clear that a calculation in the two isolated potential curves agrees well with the measured spectrum of the macrodimers at high vibrational quantum numbers.
However, for low-lying vibrational states, where the vibrational energy is comparable to the dipole-quadrupole coupling between the intersecting potentials, the vibronic coupling between them becomes relevant.
This leads to a breakdown of the Born-Oppenheimer approximation and we expect deviations from a description in the isolated potentials.
For convenience, we call the uncoupled potentials diabatic and the coupled potentials adiabatic.
In the following, we describe the calculation of the uncoupled, diabatic electronic eigenstates $|\chi^{i}_{el}(R)\rangle$ relevant for higher vibrational states from coupled, adiabatic states $|\bar{\chi}^j_{el}(R)\rangle$ obtained from the diagonalization of the Rydberg interaction Hamiltonian~\citep{Weber2017}.
In contrast to the discussion in the main text, we cannot restrict our discussion to a single potential curve, leading to additional indices $i$ and $j$ (see below).
Subsequently, we present the calculation of the exact vibronic wave functions $|\Psi^\mu_{\mathrm{Mol}}(R) \rangle$ by recoupling the obtained purely electronic diabatic eigenstates $|\chi^{i}_{el}(R)\rangle$ and purely vibrational states $\Phi_\nu(R)$ calculated in these diabatic potential wells.\\
The total atomic Hamiltonian
\begin{equation}\renewcommand\theequation{S\arabic{equation}}
\hat{H}=\hat{H}_e+\hat{T}
\end{equation}
can be split into an electronic part $\hat{H}_e = \hat{H}_1 + \hat{H}_2 + \hat{V}$ describing the spectrum of each of the atoms ($1$ and $2$) as well as the electrostatic interactions $\hat{V}$ between them, and a term $\hat{T}$ that represents the relative kinetic energy of the atom pair.
At interatomic distances $R$ larger than the LeRoy radius \cite{leroy1974}, the electrostatic interaction in the electronic part can be calculated by diagonalizing the interaction Hamiltonian obtained from a multipole expansion at various interatomic distances $R$.
The results are adiabatic electronic Born-Oppenheimer potentials $V_j(R)$, given by
\begin{equation}\renewcommand\theequation{S\arabic{equation}}
\hat{H}_{e}|\bar{\chi}^j_{el}(R)\rangle=V_j(R)|\bar{\chi}_{el}^j(R)\rangle .
\end{equation}
In position representation, the electronic eigenstates are written as $\langle{\bf r}_1,{\bf r}_2|\bar{\chi}^j_{el}(R)\rangle=\bar{\chi}^j_{el}({\bf r}_1,{\bf r}_2;R),$ where $R$ acts as a parameter.
The eigenstates $|\bar{\chi}_{el}^j(R)\rangle$ can be expanded in asymptotic pair states $\ket{\lambda}$, which are the eigenstates of the interaction in the asymptotic limit of infinite interatomic distance,
\begin{equation}\renewcommand\theequation{S\arabic{equation}}
 |\bar{\chi}^j_{el}(R)\rangle = \sum_{\lambda}\bar{c}^j_{\lambda}(R) |\lambda\rangle.
\end{equation}
The resulting adiabatic potential curves $\bar{V}_j(R)$ obtained for the macrodimer potentials relevant for this work are plotted in Fig.~\ref{supfig:5}.
In the following we calculate the diabatic states used in the discussion in the first part of the main text.
The coupling of the adiabatic eigenstates $ |\bar{\chi}^j_{el}(R)\rangle$ is provided by the kinetic energy operator $\hat{T}=-\frac{\hbar^2}{m}\partial_R^2$, where $m$ is the atomic mass.
Using the adiabatic states as a basis, we can expand the total wave function
\begin{equation}\renewcommand\theequation{S\arabic{equation}}\label{eq:psi}
 |\Psi(R)\rangle=\sum_j \bar{\Phi}_j(R)|\bar{\chi}^j_{el}(R)\rangle
\end{equation}
where $\bar{\Phi}_j(R)$ are distance-dependent amplitudes of adiabatic states $|\bar{\chi}^j_{el}(R)\rangle$ which take into acount the nuclear motion.
Substitution into the Schr\"odinger equation of the total system then yields the matrix equation~\cite{domcke_conical_2004}
\begin{equation}\renewcommand\theequation{S\arabic{equation}}
 \label{eq:schroedinger_explicit}
\left[- \frac{\hbar^2}{m}\partial_R^2+\bar{V}_{j}\right] \bar{\Phi}_{j}-\frac{\hbar^2}{m} \sum_{j^\prime}  \left[2\bar{d}_{j,j^\prime}\partial_R+  \bar{D}_{j,j^\prime}\right]\bar{\Phi}_{j^\prime}=E\bar{\Phi}_{j}
\end{equation}
where the non-adiabatic couplings are defined by
\begin{equation}\renewcommand\theequation{S\arabic{equation}}
\begin{split}
&\bar{d}_{j,j^\prime}=-\bar{d}_{j,j^\prime}=\langle \bar{\chi}^{j}_{el}(R)|\partial_R|\bar{\chi}^{j^\prime}_{el}(R) \rangle\:,\quad \\ &\bar{D}_{j,j^\prime}=\langle \bar{\chi}^j_{el}(R)|\partial_R^2|\bar{\chi}^{j^\prime}_{el}(R) \rangle.
\end{split}
\end{equation}
For isolated interaction potentials where the non-adiabatic couplings are typically negligible (as for $n=36$ in Fig.~\ref{fig:4}\,C), Eq.~(\ref{eq:schroedinger_explicit}) can be solved directly for the vibronic eigenstates and corresponding energies.
In this case, the amplitudes $\bar{\Phi}_j(R)$ can be interpreted as pure vibrational states in the adiabatic potential wells.
In our case, the non-adiabatic couplings are significant.
In order to use a shorter notation we define $\bar{\Phi}=(\bar{\Phi}_1,\bar{\Phi}_2,...,\bar{\Phi}_j,...)$ and the coupling matrices $\bar{d}$ and $\bar{D}$ with respective elements  $\bar{d}_{jj^\prime}$ and $\bar{D}_{jj^\prime}$.
Further, we introduce $\bar{V}=(\bar{V}_1,...,\bar{V}_j,...)$ and use the notation $\nabla=\mathds{1} \partial_R$ and $\Delta=\mathds{1}\partial_R^2=\nabla\nabla$.
We can write Eq.~(\ref{eq:schroedinger_explicit}) as
\begin{equation}\renewcommand\theequation{S\arabic{equation}}
\begin{split}
&\left[- \frac{\hbar^2}{m}\Delta+\bar{V}-\frac{\hbar^2}{m} \left(2d\nabla+  \bar{D}\right)\right]\bar{\Phi} \\ &=\left[- \frac{\hbar^2}{m}(\nabla+\bar{d})^{2}+\bar{V}\right]\bar{\Phi} =E\bar{\Phi},
\end{split}
\end{equation}
where we use the identity $\bar{D}=\nabla \bar{d}+\bar{d}^2$ to re-express $\Delta+2\bar{d}\nabla+\bar{D}=(\nabla+\bar{d})^2$.
With this form of the Schr\"odinger equation, the matrix $\bar{d}$ can be interpreted as a gauge potential.
Indeed we can apply a unitary transformation $U$ and obtain the same equation
\begin{equation}\renewcommand\theequation{S\arabic{equation}}
\left[- \frac{\hbar^2}{m}(\nabla+d)^2+V\right]\Phi=E\Phi
\end{equation}
with
\begin{equation}\renewcommand\theequation{S\arabic{equation}}
d=U^\dagger \bar{d} U+U^\dagger(\nabla U)\:,\:\:V=U^\dagger \bar{V} U\:,\:\:\Phi = U^\dagger\bar{\Phi}.
\end{equation}
We can now use this transformation to eliminate the additional derivative terms by choosing $U$ such that
\begin{equation}\label{cond}
\bar{d}U+\nabla U=0.
\end{equation}
In the description of the diabatic basis obtained by such a transformation, the coupling by the kinetic energy of the nuclei vanishes but $V$ becomes off-diagonal~\cite{pacher1993diabatic}.
A calculation in a purely diabatic picture can be done by neglecting the obtained off-diagonal terms.\\
For explicit calculations we can isolate two adjacent adiabatic potential curves, labeled here as $|\bar{\chi}^j_{el}(R)\rangle$ with $j\in\{1,2\}$.
In this basis we have
\begin{equation}\renewcommand\theequation{S\arabic{equation}}
\bar{d}=
\begin{bmatrix}
    0       & \bar{d}_{12}(R) \\
    -\bar{d}_{12}(R)      & 0
\end{bmatrix}
\end{equation}
and can parametrize our unitary transformation as
\begin{equation}\renewcommand\theequation{S\arabic{equation}} \label{eq:unitary_trafo}
U=
\begin{bmatrix}
    \cos \varphi(R)       & \sin \varphi(R) \\
    -\sin \varphi(R)      & \cos \varphi(R)
\end{bmatrix}.
\end{equation}
Substituting this into our Eq.~(\ref{cond}) yields
\begin{equation}\renewcommand\theequation{S\arabic{equation}}\label{eq:gauge_trafo}
0=\begin{bmatrix}
    -\sin \varphi(R)       & \cos \varphi(R) \\
    -\cos \varphi(R)      & -\sin \varphi(R)
\end{bmatrix}(\bar{d}_{12}(R)+\partial_R \varphi(R)).
\end{equation}
The gauge transformation fulfilling Eq.~(\ref{eq:gauge_trafo}) is thus given by
\begin{equation}\renewcommand\theequation{S\arabic{equation}} \label{eq:def_phi}
\varphi(R)=\int_{R_0}^R \bar{d}_{12}(R^\prime){\rm d}R^\prime
\end{equation}
where $R_0$ is chosen to be far outside of the coupled region. The two new quasi-diabatic basis states are given by
\begin{equation}\renewcommand\theequation{S\arabic{equation}} \label{eq:gauge_states}
|\chi^{i}_{el}(R)\rangle = \sum_j U_{j}^i(R)|\bar{\chi}^j_{el}(R)\rangle \equiv \sum_{\lambda} c^{i}_{\lambda}(R) |\lambda \rangle
\end{equation}
via Eq.~(\ref{eq:unitary_trafo}).
In this basis, we then obtain a two-component Schr\"odinger equation
\begin{equation}\renewcommand\theequation{S\arabic{equation}} \label{eq:schroedinger_diabatic}
\left[- \frac{\hbar^2}{m}\Delta+V\right]\Phi=E\Phi
\end{equation}
with
\begin{equation}\renewcommand\theequation{S\arabic{equation}} \label{eq:Vbar}
V=\begin{bmatrix}
    \cos^2\varphi \bar{V}_1 +\sin^2\varphi \bar{V}_2       & \cos \varphi\sin \varphi (\bar{V}_1-\bar{V}_2)\\
    \cos \varphi\sin \varphi (\bar{V}_1-\bar{V}_2)      & \sin^2\varphi \bar{V}_1 +\cos^2\varphi \bar{V}_2
\end{bmatrix}.
\end{equation}
Eq.~(\ref{eq:schroedinger_diabatic}) can be solved conveniently in two steps.
We first solve the Schr\"{o}dinger equation for each diagonal potential $V_{11}\equiv V_1(R)$ and $V_{22}\equiv V_2(R)$ numerically to obtain the set of vibrational wavefunctions $\{ \Phi^{i}_{\nu}(R) \}$ at energies $\{ E^{i}_{\nu}\}$ related with the $i^{\rm th}$ diabatic potential (see Fig.~\ref{supfig:8}\,A for the potential relevant in this work), leading to the molecular states 
\begin{equation}\renewcommand\theequation{S\arabic{equation}} \label{eq:uncoupled_state}
\ket{\Psi^{i,\nu}_{\mathrm{Mol}}(R)} = \Phi^i_\nu(R)\sum_\lambda c^i_\lambda (R)\ket{\lambda}
\end{equation}
in both diabatic potentials. Here, the nuclear motion is treated independently from the electronic interaction petential in a Born-Oppenheimer framework. Now, the residual coupling due to the off-diagonal terms is then accounted for by diagonalizing the matrix $V$ in the basis of the $\Phi^{i}_{\nu}(R)$, which gives the vibronically coupled eigenstates (see Fig.~\ref{supfig:8}\,B)
\begin{equation}\renewcommand\theequation{S\arabic{equation}} \label{eq:recoupled_states}
| \Psi^\mu_{\mathrm{Mol}}(R)  \rangle = \sum_{i\nu} C^{\nu}_{i\mu} \Phi^{i}_{\nu}(R)|\chi^{i}_{el}(R)\rangle.
\end{equation}
Inserting  Eq.~(\ref{eq:gauge_states}) into Eq.~(\ref{eq:recoupled_states}), the vibronically coupled eigenstates can then be re-expressed as
\begin{equation}\renewcommand\theequation{S\arabic{equation}}\label{eq:recoupled_states2}
 |\Psi^\mu_{\mathrm{Mol}}(R) \rangle = \sum_{i\nu} C^{\nu}_{i\mu} \Phi^{i}_{\nu}(R)\sum_{\lambda} c^{i}_{\lambda}(R) |\lambda \rangle \equiv \sum_{\lambda} \tilde{c}^{\mu}_{\lambda}(R) |\lambda \rangle,
\end{equation}
with $\tilde{c}^{\mu}_{\lambda}(R) = \sum_{i\nu}C_{\nu}^{i\mu} \Phi_{i}^{\nu}(R)c^{i}_{\lambda}(R)$.
In the following calculations of the optical coupling to the macrodimers $\ket{\Psi^{\nu}_{\mathrm{Mol}}(R)}$ in our $0^+_g$ potential, we omit the index $i$ in Eq.~(\ref{eq:uncoupled_state}) because a diabatic description is only relevant for the optically coupled eigenstates in the potential $V_1(R)$, see Fig.~\ref{supfig:8}\,A.
\section{Optical coupling}
\begin{figure}[htp]
\renewcommand\thefigure{S\arabic{figure}}
 \renewcommand\theHfigure{Supplement3./thefigure}
  \centering
  \includegraphics{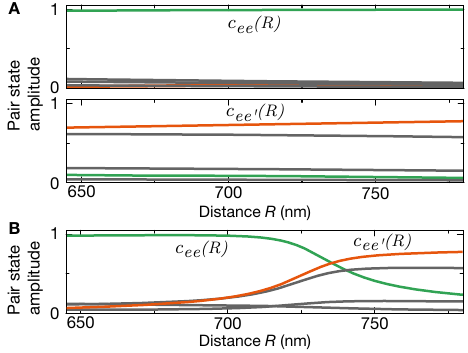}%
  \caption{\label{supfig:9}
 \textbf{State decomposition into asymptotic pair states.}
 (\textbf{A}) State decomposition of $\ket{\widetilde\alpha(R)}$ (upper panel) and $\ket{\widetilde\beta(R)}$ (lower panel), which are dominated by the states $\ket{e,e}$ (green) and $\ket{e,e^\prime}$ (orange).
The weak spatial dependence of the amplitudes justifies the approximation of constant couplings made in the main text. Gray lines are other, weaker contributing pair states with strongly suppressed or absent optical couplings.
 (\textbf{B}) State decomposition of the coupled electronic state $\ket{\chi_{el} (R)} = c_{\widetilde \alpha}(R)\ket{\widetilde\alpha(R)}+c_{\widetilde \beta}(R)\ket{\widetilde\beta(R)}$ for the binding potential $V_1(R)$. We find that also the approximation $\ket{\chi_{el} (R)} \approx c_{ee}(R)\ket{e,e}+c_{ee^\prime}(R)\ket{e,e^\prime}$ captures all relevant details to understand the optical coupling mechanism (see Eq.~(\ref{eq:Omega_par_approx}), Eq.~(\ref{eq:Omega_orth_approx}) and Table~\ref{Table:2}). For our results presented in Fig.~\ref{supfig:7} we used neither of both approximations. 
}
\end{figure}

In the main text, we described the macrodimer potential by the three crossing van-der-Waals states $\ket{\widetilde\alpha(R)}$, $\ket{\widetilde\beta(R)}$ and $\ket{\widetilde\gamma(R)}$. 
The weak dependence on the interatomic distance due to dispersive van-der-Waals interactions between off-resonant asymptotic pair states was neglected in the main text.
Due to this weak dependence, the two-photon Rabi couplings $\widetilde\Omega_{\widetilde\alpha}(R)(\widetilde\Omega_{\widetilde\beta}(R))$ are sligthly spatial dependent because of the changing state admixture in $\ket{\widetilde\alpha(R)}(\ket{\widetilde\beta(R)})$.
Here, we use the expansions Eq.~(\ref{eq:uncoupled_state}) and Eq.~(\ref{eq:recoupled_states2}) into asymptotic pair states instead.
These asymptotic pair states are the eigenstates in the non-interacting limit of large separations. 
Since it is straightforward to expand the asymptotic states into products of single-particle states, optical coupling rates to these pair states are easy to calculate. 
As we will see in Eq.~(\ref{eq:seperationF_C_O}), this allows for a strict separation of spatially independent optical couplings to contributing pair states $\ket{\lambda}$ and generalized Franck-Condon overlap integrals taking into account the spatial dependent amplitudes of these pair states. \\
The two dominating asymptotic pair states contributing to our $0^+_g$ potential are the non-interacting pair states $\ket{e,e}$ and $\ket{e,e^\prime}$ with coefficients $c_{ee}(R)$ and $c_{ee^\prime}(R)$, which are the asymptotic limits of the van-der-Waals states $\ket{\widetilde \alpha (R)}$ and $\ket{\widetilde \beta (R)}$. In the main text, these asymptotic states were defined via the states $\ket{e} = \ket{35P_{1/2}}$ and $\ket{e^\prime} = \ket{35P_{3/2}}$. Accounting also for the spin projections of both atoms and symmetrizing the states with respect to the inversion and reflection symmetry of the $0^+_g$~\citep{Weber2017} potential leads to
\begin{equation}\renewcommand\theequation{S\arabic{equation}} \label{eq:onetilde}
\ket{e,e}=1/\sqrt{2}\left(\ket{e\uparrow,e\downarrow}-\ket{e\downarrow,e\uparrow}\right),
\end{equation}
\begin{equation}\renewcommand\theequation{S\arabic{equation}}\label{eq:twotilde}
\begin{split}
\ket{e,e^\prime} \hspace{2pt}=\hspace{4pt}  1/2( & \ket{e\uparrow,e^\prime\downarrow} -\ket{e^\prime\downarrow,e\uparrow} \\ + & \ket{e\downarrow,e^\prime\uparrow} - \ket{e^\prime\uparrow,e\downarrow}),
\end{split}
\end{equation}
with spin orientations $\ket{\uparrow}=\ket{m_J=+1/2}$ and $\ket{\downarrow}=\ket{m_J=-1/2}$ relative to the molecular axis. A distance dependent decomposition of the states $\ket{\widetilde \alpha(R)}$ and $\ket{\widetilde \beta(R)}$ which form the potential well $V_1(R)$ is shown in Fig.~\ref{supfig:9}.
Among other states, we find an optically uncoupled pair state including single particle states $35S_{1/2}$ and $36S_{1/2}$ and weaker admixed pair states, where both atoms are in $\ket{e^\prime}$.
The optical coupling to the latter are suppressed not only by smaller pair state amplitudes, but also because they are only accessible via the intermediate state $\ket{e^\prime}$. Due to the fine-structure splitting of $2566$\,MHz for our Rydberg state, ten times larger compared to the typical detunings to $\ket{e}$, we restrict ourself to the intermediate pair states states including one atom in the ground state and a second atom in $\ket{e\uparrow}$ or $\ket{e\downarrow}$ in the following calculations.  
\subsection{Molecular symmetry}
\begin{figure}[htp]
\renewcommand\thefigure{S\arabic{figure}}
 \renewcommand\theHfigure{Supplement4./thefigure}
  \centering
  \includegraphics{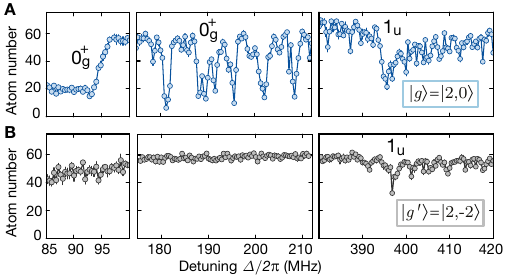}%
  \caption{\label{supfig:2}
  \textbf{Excitation control by pair state symmetry.}
  (\textbf{A}) Starting from all atoms in $\ket{g}$, we can couple to both molecular potentials $0^+_g$ and $1_u$.
  Due to the smaller detunings in the left panel, we chose a shorter illumination time of $10\,$ms to avoid saturation.
  The other panels correspond to a zoom into Fig.~\ref{fig:2}\,A.
  (\textbf{B}) Starting from $\ket{g^\prime}=\ket{5S_{1/2},\,F=2,\,m_F=-2}$, we can only couple to $1_u$ because the ground state is a product state of electronic and nuclear angular momenta.
  While the illumination time was $20\,$ms in the left panel, the other measurements were taken at $100\,$ms.
  All error bars denote one s.e.m.
  }
\end{figure}

This work mainly discusses the macrodimers in the $0^+_g$ potential with gerade symmetry and zero angular momentum projection on the intermolecular axis.
Interestingly, the antisymmetric states (see definition of $\ket{e,e}$ and $\ket{e,e^\prime}$ in Eq.~(\ref{eq:onetilde}) and Eq.~(\ref{eq:twotilde})) in the $0^+_g$ potential can only be optically coupled because of the hyperfine coupling of the ground state, which is negligible for Rydberg states.
This is due to the fact that the initial pair states are symmetric and that the optical coupling preserves the pair state symmetry.
Rewriting the initial pair state $\ket{g,g}$ with $\ket{g}=\ket{5S_{1/2},\,F=2,\,m_F=0}$ into the components of the electronic angular momentum $J$ and the nuclear spin $I$ leads to
\begin{equation}\renewcommand\theequation{S\arabic{equation}} \label{eq:Initial_pair}
\ket{g,g} = \sqrt{3}/2\ket{+}_J\ket{+}_I + 1/2\ket{\unaryminus}_J\ket{-}_I.
\end{equation}
Here, only the antisymmetric part $\ket{\unaryminus}_{J} = 1/\sqrt{2}(\ket{5S_{1/2}\uparrow,5S_{1/2}\downarrow}-\ket{5S_{1/2}\downarrow,5S_{1/2}\uparrow})$ can couple to the desired potential.
We probe this experimentally by starting with the initial pair state $\ket{g^\prime,g^\prime}$ with $\ket{g^\prime}=\ket{5S_{1/2},\,F=2,\,m_F=-2}$ which does not contain antisymmetric fine-structure components, as shown in Fig.~\ref{supfig:2}.
All other experimental parameters are identical to the spectroscopy with initial state $\ket{g,g}$.
As expected, we find spectroscopic features for the lower and the upper branch of the avoided crossing in the $0^+ _g$ potential only by starting from $\ket{g}$, while the symmetric $1_u$ potential can be observed for both initial states.
Eq.~(\ref{eq:Initial_pair}) also implies that our photoassociation procedure projects both nuclear spins into the entangled state $\ket{-}_I=1/\sqrt{2}(\ket{3/2\uparrow,3/2\downarrow}-\ket{3/2\downarrow,3/2\uparrow})$.
\subsection{Calculation of Rabi frequencies}
The Hamiltonian describing the optical coupling is given by
\begin{equation}\renewcommand\theequation{S\arabic{equation}}
 \hat{H}_L = - e \mathbf{E} \cdot(\mathbf{ \hat{r}}_1 + \mathbf{ \hat{r}}_2)
\end{equation}
where $\mathbf{ \hat{r}}_1$ and $\mathbf{ \hat{r}}_1$ are the position operators of both electrons and $\mathbf{E}$ is the electric field of the excitation light.
For the calculation of the optical couplings to the macrodimers~\citep{Samboy2011,Optical_coupling_macrodimers} we have to account for the fact that the eigenstates of the interatomic interaction are only symmetric with respect to rotations around the molecular axis but not necessarily spherical symmetric.
Therefore, the optical couplings to pair states $\ket{\lambda}$ depend on the relative orientation of the polarization of the light and the molecular axis.
Hence, we have to rotate our initial state $\ket{g,g}$, which is defined with respect to the initial quantization axis of the magnetic field to be parallel to the quantization axis of the molecule where the states $\ket{\lambda}$ are defined.
Here, we make use of the decomposition~(\ref{eq:Initial_pair}) into optically uncoupled symmetric and coupled antisymmetric initial pair states where $\ket{\unaryminus}_J$ is formally equivalent to a rotationally symmetric spin singlet state.
This allows us to use $\ket{\unaryminus}_J$ as the initial state for all molecular orientations and the only parameter which depends on the interatomic axis is the polarization of the excitation light.\\
\begin{figure*}
\renewcommand\thefigure{S\arabic{figure}}
 \renewcommand\theHfigure{Supplement5./thefigure}
  \centering
  \includegraphics{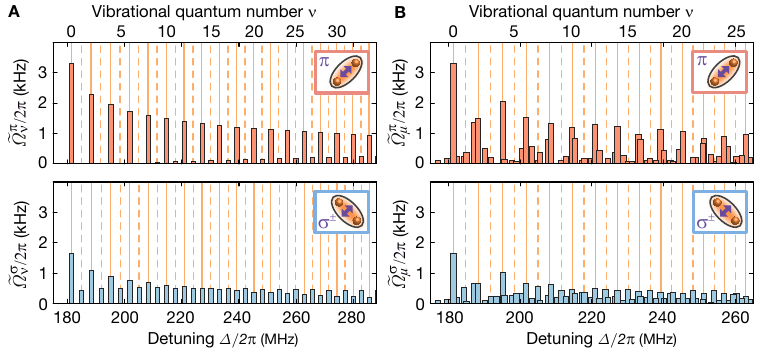}%
  \caption{\label{supfig:7}
  \textbf{Calculated Rabi coupling to macrodimer resonances.} 
  (\textbf{A}) Optical coupling to molecules $\ket{\Psi^\nu_{\textrm{Mol}}(R)}$ in the diabatic potential $V_1(R)$ for the two possible spatial configurations.
For $\widetilde\Omega^{\pi}_{\nu}$ (red), we recognize strong coupling to even vibrational states while the odd ones are suppressed.
 A comparison of the relative strength at the individual resonances yields $\widetilde\Omega^{\pi}_{\nu}/\widetilde\Omega^{\sigma}_{\nu}>1$ for even $\nu$ and $\widetilde\Omega^{\sigma}_{\nu}/\widetilde\Omega^{\pi}_{\nu}>1$ for odd $\nu$, which explains the observed orientation of the photo-associated molecules. 
At higher vibrational quantum numbers and $\sigma^\pm$-polarization, we additionally find that odd resonances are stronger coupled than even resonances, see also Fig.~\ref{supfig:1}\,B. 
  (\textbf{B}) Optical coupling to the vibronically coupled states $|\Psi^\mu_{\mathrm{Mol}}(R) \rangle$ shown in Fig.~\ref{supfig:8}\,B.
Because many of the additional resonances originating from the intersecting potential well are only weakly coupled by the light field, we only show states with significant coupling in Fig.~\ref{fig:4}\,B.
  Orange solid/dashed lines indicate the calculated even/odd vibrational states in the diabatic potential $V_1(R)$.}
\end{figure*}
In the case of a polarization parallel to the molecular axis, the light is $\pi$-polarized and the only relevant intermediate states which can couple to the $0^+_g$ potential (see Eq.~(\ref{eq:onetilde}) and Eq.~(\ref{eq:twotilde})) are antisymmetric states $\ket{i^\pi_1} =  1/\sqrt{2}(\ket{5S_{1/2}\uparrow,e\downarrow}-\ket{5S_{1/2}\downarrow,e\uparrow})$ and $\ket{i_2^\pi} = 1/\sqrt{2}(\ket{5S_{1/2}\downarrow,e\uparrow} - \ket{e\uparrow,5S_{1/2}\downarrow})$. 
The coupling strength $\Omega^{\pi}_{1}$ from $\ket{g,g}$ to the intermediate state is given by
\begin{equation}\renewcommand\theequation{S\arabic{equation}}
\Omega^{\pi}_{1} = \bra{i^\pi_{1,2}}\hat{H}_L\ket{g,g}  = 1/2\bra{i^\pi_{1,2}}\hat{H}_L\ket{\unaryminus}_J,
\end{equation}
where we used that the coupling elements to both intermediate states turns out to be the same. 
For the coupling $\Omega^{\pi}_{2}$ from the intermediate states to the macrodimer states, we have to take into account the different pair states $\ket{\lambda}$ contributing to Eq.~(\ref{eq:uncoupled_state}). 
Additionally, generalized Franck-Condon integrals $f^\nu_{\lambda} = \int\Phi_\nu^{\,*}(R)c^*_{\lambda}(R)\Phi_{g}(R){\rm d}R$ now appear, with the initial relative wave function of two nuclei in the lattice~\cite{BlochZwergerDalibard}
\begin{equation}\renewcommand\theequation{S\arabic{equation}}
\Phi_g(R)=1/(2\pi\sigma_{\text{lat}}^2)^{1/4}e^{-(R-R_{\textrm{in}})^2/(4\sigma_{\text{lat}}^2)}.
\end{equation}
Here, $\sigma_{\text{lat}}$ is the extension of the on-site ground state wave function for a single atom and $R_{\textrm{in}}$ is the diagonal distance in the lattice. 
This leads to
\begin{equation}\renewcommand\theequation{S\arabic{equation}}
\begin{split}
\label{eq:seperationF_C_O}
\Omega^{\pi}_{2} & = \int \Phi_\nu^{\,*}(R)\sum_\lambda c^*_\lambda(R)\bra{\lambda}\hat{H}_L \ket{i^\pi_{1,2}}\Phi_g(R){\rm d}R \\
& = \sum_\lambda\bra{\lambda}\hat{H}_L \ket{i^\pi_{1,2}}f^\nu_{\lambda} = \sum_\lambda\Omega^\pi_{\lambda}f^\nu_{\lambda} ,
\end{split}
\end{equation}
where we introduced the spatially independent optical couplings $\Omega^\pi_{\lambda} = \bra{\lambda}\hat{H}_L \ket{i^\pi_{1,2}}$ to the contributing non-interacting pair states $\ket{\lambda}$. 
Applying a rotating-wave approximation in a frame co-rotating with the laser frequency, the full coupling Hamiltonian, in the basis $\{\ket{g,g},\ket{i^\pi_{1}},\ket{i^\pi_{2}},\ket{\Psi^{\nu}_{\textrm{Mol}}(R)}\}$, reads
\begin{equation}\renewcommand\theequation{S\arabic{equation}}\label{eq:Hamiltoninan_Matrix}
\hat{H}_L = \hbar\begin{pmatrix}
0 \hspace{501pt}& \Omega^{\pi}_{1}/2 & \Omega^{\pi}_{1}/2 & 0 \\  \Omega_{1}^{\pi*}/2 & -\Delta & 0 & \Omega^{\pi}_{2}/2 \\
\Omega_{1}^{\pi*}/2 & 0 & -\Delta & \Omega^\pi_{2}/2 \\ 0 & \Omega_{2}^{\pi*}/2 & \Omega_{2}^{\pi*}/2 & 0
\end{pmatrix},
\end{equation}
if the laser is two-photon resonant (i.e. $E_{\nu} = 2\Delta$) to a macrodimer state $\ket{\Psi^{\nu}_{\textrm{Mol}}(R)}$. Since $\Delta\gg\Omega$, the two intermediate states can be adiabatically eliminated, leading to a total effective Rabi coupling from $\ket{g,g}$ to $\ket{\Psi^{\nu}_{\textrm{Mol}}(R)}$
\begin{equation}\renewcommand\theequation{S\arabic{equation}}\label{eq:Omega_par}
\widetilde\Omega^{\pi}_{\nu} = \frac{\Omega^\pi_{1}\Omega^{\pi}_{2}}{\Delta}=\frac{\Omega^{\pi}_{1}}{\Delta}\sum_\lambda \Omega^{\pi}_{\lambda} f^\nu_{\lambda}.
\end{equation}
\begin{table}[htp]
\renewcommand\thetable{S\arabic{table}}
  \begin{tabular}{ | l | c | c | c | c | r |}
  \hline
    \raisebox{10pt}
        &   $f^\nu_{ee}/f^\nu_{ee^\prime}$ & $\Omega^\pi_{ee}/\Omega^\pi_{ee^\prime}$ & $\Omega^\sigma_{ee}/\Omega^\sigma_{ee^\prime}$ & $\widetilde \Omega^\pi_\nu$ & $\widetilde \Omega^\sigma_\nu$ \\ \hline
        even $\nu$ & + & + & -  & strong & weak \\ \hline
       odd $\nu$ & -  & + & -   & weak   & strong \\ \hline
  \end{tabular}
  \caption{\label{Table:2}
 \textbf{Qualitative understanding of the optical coupling.} 
The relative wavepacket $\Phi_g(R)$ is only slowly varying over the extension of the vibrational states $\Phi_\nu(R)$ and the dominating coefficients $c_{ee}(R)$ and $c_{ee^\prime}(R)$ in the electronic state $\ket{\chi_{el}(R)}$ swap roles at the potential minimum, see Fig.~\ref{fig:1}\,B and Fig.~\ref{supfig:9}. As a consequence, $f^\nu_{ee}$ and $f^\nu_{ee^\prime}$ are essentially measuring weights of the vibrational wave function $\Phi_{\nu}(R)$ at both sides of the potential. Therefore, the relative sign of $f^\nu_{ee}$ and $f^\nu_{ee^\prime}$ depends on the parity of the vibrational state. Additionally, the relative sign of the couplings $\Omega_{ee}$ and $\Omega_{ee^\prime}$ depends on the UV polarization relative to the molecular axis.}
\end{table}
Within a simplified description where we only account for the dominating asymptotic states $\ket{e,e}$ and $\ket{e,e^\prime}$, the sum reduces to
\begin{equation}\renewcommand\theequation{S\arabic{equation}}\label{eq:Omega_par_approx}
\begin{split}
\widetilde\Omega^\pi_\nu &\approx \frac{\Omega^\pi_{1}}{\Delta}(\Omega^\pi_{ee}f^\nu_{ee} + \Omega^\pi_{ee^\prime}f^\nu_{ee^\prime} ) \\ &= \frac{\Omega^2}{\Delta}(f^\nu_{ee}+1.3f^\nu_{ee^\prime}),
\end{split}
\end{equation}
with the experimentally calibrated single-photon coupling rate $\Omega/2\pi=1.2(1)\,$MHz and a numerical factor $1.3$ taking into account slightly different reduced matrix elements between the ground state $\ket{g}$ and the two fine-structure states $\ket{e}$ and $\ket{e^\prime}$.\\
In the case of a polarization perpendicular to the molecular axis, the light is $\sigma^\pm$-polarized and the intermediate states used above cannot be coupled. 
Now, we find $\ket{i^{\sigma}_1} =  1/\sqrt{2}(\ket{5S_{1/2}\uparrow,e\uparrow}-\ket{5S_{1/2}\uparrow,e\uparrow})$ and  $\ket{i^{\sigma}_2} = 1/\sqrt{2}(\ket{5S_{1/2}\downarrow,e\downarrow} - \ket{e\downarrow,5S_{1/2}\downarrow})$.
An analogous theoretical analysis leads to
\begin{equation}\renewcommand\theequation{S\arabic{equation}}\label{eq:Omega_orth}
\widetilde\Omega^{\sigma}_{\nu} = \frac{\Omega^{\sigma}_{1}\Omega^{\sigma}_{2}}{\Delta}=\frac{\Omega^{\sigma}_{1}}{\Delta}\sum_\lambda \Omega^{\sigma}_{\lambda} f^\nu_{\lambda},
\end{equation}
which again can be approximated by
\begin{equation}\renewcommand\theequation{S\arabic{equation}}
\label{eq:Omega_orth_approx}
\begin{split}
\widetilde\Omega^{\sigma}_{\nu} & \approx \frac{\Omega^\pi_{1}}{\Delta}(\Omega^\sigma_{ee}f^\nu_{ee} + \Omega^\sigma_{ee^\prime}f^\nu_{ee^\prime})\\ &= \frac{\Omega^2}{\Delta}(f^\nu_{ee}-\frac{1.3}{2}f^\nu_{ee^\prime}).
\end{split}
\end{equation}
Comparing both approximated equations, one can recognize a flip in the relative sign between the optical couplings to $\ket{e,e}$ and $\ket{e,e^\prime}$, which is the origin of the alternating molecular orientation, see Table~\ref{Table:2}.
It turns out that Eq.~(\ref{eq:Omega_par_approx}) and Eq.~(\ref{eq:Omega_orth_approx}) capture all essential physics to understand the optical coupling mechanism and only slightly deviates from a more accurate calculation.
\begin{figure}[htp]
\renewcommand\thefigure{S\arabic{figure}}
 \renewcommand\theHfigure{Supplement6./thefigure}
  \centering
  \includegraphics{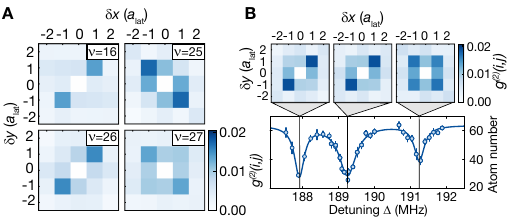}%
  \caption{\label{supfig:3}
  \textbf{Additional correlation measurements}.
  (\textbf{A}) The alternating molecular orientation for even and odd vibrational states can also be observed for higher $\nu$.
  (\textbf{B}) For the modified spectrum around the gap close to the potential miminum, the picture of alternating even and odd parity vibrational wave functions, however, breaks down (see also Fig.~\ref{supfig:7}\,B).
  All error bars denote one s.e.m.
  }
\end{figure}
\begin{figure*}[htp]
\renewcommand\thefigure{S\arabic{figure}}
 \renewcommand\theHfigure{Supplement7./thefigure}
  \centering
  \includegraphics{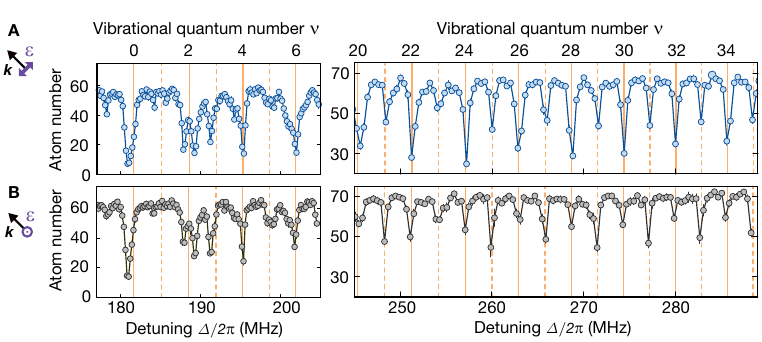}%
  \caption{\label{supfig:1}
  \textbf{Polarization dependence of the atom loss spectrum} 
  (\textbf{A}) For a laser polarization parallel to the atomic plane, even lines dominate the loss spectrum.
  (\textbf{B}) For an excitation laser polarized perpendicular to the atomic plane, in agreement with the calculations, this changes for higher vibrational states. 
Besides this figure and Fig.~\ref{fig:3}\,B, the excitation light was always polarized in the atomic plane.
Dashed orange lines represent again the calculated resonances in the diabatic potential well and all error bars denote one s.e.m.
}
\end{figure*}This is not only because of the strong domination of the states $\ket{e,e}$ and $\ket{e,e^\prime}$ in the pair potentials but also because other contributing pair states experience absent or strongly supressed optical couplings. \\
A rigorous calculation using Eq.~(\ref{eq:Omega_par}) and Eq.~(\ref{eq:Omega_orth}) and taking into account a large set of basis states as well as the experimentally calibrated Rabi frequency $\Omega/2\pi = 1.2(1)\,$ MHz is shown in Fig.~\ref{supfig:7}\,A.
While the discussion so far was focused on states $|\Psi^{\nu}_{\textrm{Mol}}(R)\rangle$ in the diabatic potential $V_1(R)$, the optical couplings $\widetilde\Omega^{\pi}_{\mu}$ and $\widetilde\Omega^{\sigma}_{\mu}$ to the vibronically coupled states $| \Psi^{\mu}_{\mathrm{Mol}}(R)  \rangle$ can be calculated by replacing the pair state amplitudes and vibrational wave functions in the generalized Franck-Condon integrals by the coefficients $\tilde{c}^{\mu}_{\lambda}(R)$ defined in Eq.~(\ref{eq:recoupled_states2}).
The result of this calculation is shown in Fig.~\ref{supfig:7}\,B.
Besides a splitting of the eigenergies in the diabatic potential due to the vibronic coupling, the coarse structure and the polarization dependence remain qualitatively the same.
\subsection{Further experimental tests}
\begin{figure}[htp]
\renewcommand\thefigure{S\arabic{figure}}
 \renewcommand\theHfigure{Supplement8./thefigure}
  \centering
  \includegraphics{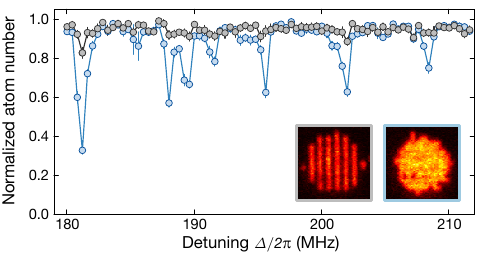}%
  \caption{\label{supfig:4}
  \textbf{Excitation control by density modulation.} 
  Changing the initial density distribution such that there are no atom pairs at distances of $\sqrt{2}\alat$ leads to a strong suppression of the macrodimer resonances. 
  For both spatial configurations, the atom number was normalized to the initial atom number without UV light and the illumination time was $t_{\textrm{UV}}=20\,$ms. 
  All error bars denote one s.e.m.
  }
\end{figure}

In Fig.~\ref{supfig:3}\,A, we show additional correlation measurements with the same behaviour as presented in the main text for higher vibrational quantum numbers up to $\nu = 27$.
Interestingly, the concept of even and odd vibrational wave functions breaks down around the intersection (see Fig.~\ref{supfig:3}\,B) where the splitting of the lines due to the vibronic coupling does not allow a classification in alternating even and odd vibrational quantum numbers.
Measured correlations at the three resonances around $\Delta/2\pi \approx 190\,$ MHz show that the two left resonances feature almost the same directionality in the correlations, while the right resonance at $\Delta/2\pi = 192\,$MHz features almost isotropic correlations.\\   
As briefly mentioned in the main text, rotating the UV polarization out of plane modifies the relative strength of even and odd macrodimer resonances, see Fig.~\ref{supfig:1}.
 This can be understood from the microscopic picture developed above.
For the configuration shown in Fig.~\ref{supfig:1}\,A, molecule formation occurs due to coupling rates $\widetilde\Omega^{\pi}_{\nu}$ as well as $\widetilde\Omega^{\sigma}_{\nu}$.
 Because of the strong coupling $\widetilde\Omega^{\pi}_{\nu}$ for even vibrational states, these states dominate the total loss spectrum, see Fig.~\ref{supfig:7}.
 Additionally, as predicted and verified in the correlation measurements, losses at even(odd) resonances occur mainly due to molecule formation parallel(perpendicular) to the polarization of the UV light.
In the case of Fig.~\ref{supfig:1}\,B, both molecular orientations are orthogonal to the dimer axis and we only probe $\widetilde\Omega^{\sigma}_{\nu}$. 
We find that even resonances are weaker but still dominating for low vibrational states. 
However, consistent with the calculations shown in the lower panels of Fig.~\ref{supfig:7}, this flips for higher vibrational states where the odd resonances become stronger. \\
Furthermore, we probed the strong distance dependence of the losses by changing the spatial configuration of the atoms in the lattice.
We use our ability to locally remove atoms from our initial Mott insulator by optically adressing individual lattice sites with our microscope~\cite{weitenberg11} to create an initial pattern of alternatingly populated and empty rows in the optical lattice for which atom pairs at a distance of $\sqrt{2}\alat$ are absent.
For both initial configurations, we probe the first macrodimer resonances spectroscopically, see Fig.~\ref{supfig:4}.
As expected, the formation of macrodimers is strongly suppressed in the new density modulated configuration.
A closer look reveals that the atom loss at the macrodimer resonances is still larger compared to the background loss due to the off-resonant coupling to the bare Rydberg resonance.
Measuring decay constants for both cases shows that we can suppress the decay by roughly one order of magnitude by density modulating the initial Mott insulator.
The remaining coupling is consistent with the macrodimer excitation of two atoms at neighboring lattice sites, where the Franck-Condon factor is much smaller but not zero.
\section{Experimental details}
\begin{figure}[htp]
\renewcommand\thefigure{S\arabic{figure}}
 \renewcommand\theHfigure{Supplement9./thefigure}
  \centering
  \includegraphics{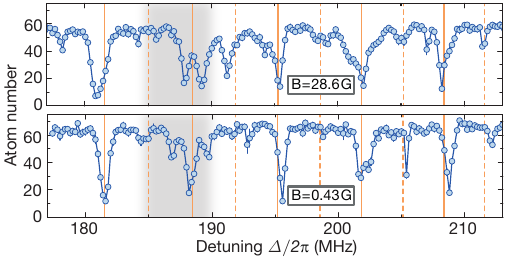}%
  \caption{\label{supfig:10}
  \textbf{Magnetic field dependence and vibronic coupling.}
 Signal at $B = 28.6\,$G (perpendicular to the atomic plane) and at $B = 0.43 \,$G (parallel to the wavevector of the UV light). The main effect is a modification of the line splitting due to the intersection at low quantum numbers (gray shaded region). The orange lines indicate the unsplit energies of the diabatic eigenstates. For both cases, the detuning $\Delta$ was measured relative to the center of the two Zeeman sublevels. All error bars denote one s.e.m.
}
\end{figure}

For most of the data presented in this work (also for $n=36$), we applied a magnetic field $B = 28.6\,$G perpendicular to the atomic plane.
 At this field strength, the Rydberg state $\ket{e}$ splits into two Zeeman sublevels separated by $\pm13.3\,$MHz. 
The magnetic field is expected to be not a critical parameter for our $0^+_g$ potential with zero angular momentum projection, even if the intermediate states $\ket{e}$ get Zeeman splitted. 
This was verified by comparing the spectroscopic signal at low vibrational quantum numbers for two different field values, see Fig.~\ref{supfig:10}.
As expected, the energy of the diabatic eigenstates stays mainly unaffected. 
However, we find that the splitting of the lines with respect to the diabatic eigenergies due the intersecting potential gets modified.
This is also expected to be more sensitive because it critically depends on the relative detuning of the vibrational modes in both diabatic potentials $V_1(R)$ and $V_2(R)$.
Besides Fig.~\ref{supfig:10}, the spectroscopic data for low field was also used in Fig.~\ref{fig:4}\,B where we discuss the breackdown of the Born-Oppenheimer approximation around the intersection. 
Because a numerical treatment of the vibronic coupling at finite magnetic field is numerically too challenging due to the additional Zeeman sublevels appearing in the potential calculations, we reduced the field to $B = 0.43\,$G.
 All theoretical calculations in this work were done at zero field where the number of contributing pair states can be significantly increased.\\
During the experiment, the atoms were pinned in the lattice at lattice depths of $40E_\textrm{r}$ for both lattice directions in the atomic plane, where $E_\textrm{r} = h^2/(8m a_{\textrm{lat}}^2)$ is the recoil energy. For the lattice perpendicular to the atomic plane, we chose a depth of $80E_\textrm{r}$.
If not stated otherwise, the UV light was polarized in the atomic plane.
For all measurements in this work, for $n=35,36$ and both polarizations of the UV light, we used a single-photon Rabi coupling of $\Omega/2\pi = 1.2(1)\,$MHz between the ground state and the Rydberg state.
In the case of the ground state $\ket{g^\prime}$ discussed in Fig.~\ref{supfig:2}, the laser power was reduced in order to compensate a larger Clebsch-Gordan coefficient.
As in our previous works~\cite{zeiher_many-body_2016}, $\Omega$ was calibrated by Ramsey spectroscopy in the ground state manifold.
The detuning $\Delta$ was always measured relative to the center of the two Zeeman sublevels.
\subsection{Spectroscopic data}
The broad spectrum detuned from the Rydberg resonance shown in Fig.~\ref{fig:2}\,A contains roughly $1500$ data points with frequency spacing of $400\,$kHz.
Each datapoint represents the average atom number of around 10 experimental shots, analyzed in a circular region-of-interest of 80 sites around the center of the cloud.
In order to capture all narrow-linewidth macrodimer resonances, the frequency was swept over a range of $\pm 240\,$kHz during the illumination time of $t_{\textrm{UV}}=100\,$ms.
The two isolated resonances at the red-detuned side of the spectrum are technical artefacts due to the relative detuning of $230.8\,$MHz of the two beams creating our in-plane optical lattice.
\begin{table*}[htp]
\renewcommand\thetable{S\arabic{table}}
  \begin{tabular}{ | l | c | c | c | c | c | c | c | r |}
    \hline     
    $\textrm{Vib. state } \nu$ \raisebox{9pt}    & 0 & 9 & 12 & 13 & 16 & 25 & 26 & 27  \\ \hline 
        $\textrm{Time } t_{\textrm{UV}} \textrm{ (ms)}$ \raisebox{9pt} & 0.12 & 5.0 & 1.5  & 10  & 0.9  & 10  & 3.5  & 3  \\ \hline     
            $\textrm{Exp. shots}$ \raisebox{9pt}& 158 & 240 & 571  & 283  & 197  & 274  & 256  & 264   \\  \hline    
                $\textrm{Filling}$ \raisebox{9pt}& 87.7(5) & 88.4(3) & 87.6(3)  & 87.4(5)  & 89.7(4)  & 85.1(4)  & 86.5(5)  & 86.6(4)  \\ \hline 
    $\gtwo(1,1)\times10^{-2}$ \raisebox{9pt}    & 1.90(17) & 0.64(10) & 2.05(08)  & 0.49(10)  & 1.39(12)  & 0.61(11)  & 1.53(11)  &  1.01(10)   \\ \hline   
    $\gtwo(1,\unaryminus 1)\times10^{-2}$  \raisebox{9pt}   & 0.95(14) & 1.51(13) & 0.72(06)  & 1.45(11)  & 0.59(11)  & 1.74(13)  & 0.52(10)  & 1.35(12)\\     
 \hline
  \end{tabular}
  \caption{\label{Table:1}
  \textbf{Details on the correlation measurements.} 
The measured ratios of the correlations $\gtwo(1,1)$ and $\gtwo(1,-1)$ along both lattice diagonals are smaller than the theoretical expected values for even and odd lines, see Fig.~\ref{supfig:7} and note that scattering rates are proportional to the square of the Rabi coupling. We attribute this to a small positive background in the correlations which shifts the measured directionality in the excitation rate towards a more balanced value. Experimental errors on the correlations are calculated by using a delete-1 Jackknife algorithm, errors on the filling denote one s.e.m.
}
\end{table*}As a result, the Zeeman-splitted transitions from $\ket{g}$ to $\ket{e\uparrow}$ and $\ket{e\downarrow}$ reappear on both sides of the Rydberg resonance.
While the sidebands at the red-detuned side of the resonance are recognized easily, they are overlapped with the macrodimer resonances $\nu=11$ and $\nu=20$ on the blue-detuned side.
The overall decrease of the coupling strength for even vibrational quantum numbers observed in Fig.~\ref{fig:2}\,A and  Fig.~\ref{fig:2}\,B is smaller compared to the actual value due to the strong saturation of the lower macrodimer lines.
For the high-resolution spectroscopy of the lowest vibrational level shown in Fig.~\ref{fig:2}\,C, we illuminate the cloud only for $t_{\textrm{UV}}=0.8\,$ms in order to avoid artificial broadening.
Additionally, we did not sweep the UV-frequency because the width of the resonance is larger than the separation of the data points.
We also repeated the spectroscopy of the lowest line for lower powers and verified that there is no power broadening.
This is not surprising because $\widetilde \Omega_\nu$ is smaller than the decay rate which is expected to be at least twice as large as the decay rate $\Gamma_{e}\approx 25\,\textrm{ms}^{-1}$ of an isolated Rydberg atom in $\ket{e}$.
In Fig.~\ref{fig:2}\,C, one can also recognize a region where the Lorentzian fit deviates slightly from the data.
This is reproducible and most likely due to a weakly coupled state originated from the intersecting state $\ket{\widetilde \gamma(R)}$, which overlaps with the first strong macrodimer line.
For the additional spectroscopic data shown in Fig.~\ref{fig:4}, Fig.~\ref{supfig:10} and the left panels of Fig.~\ref{supfig:1}, different data points are separated by $200$\,kHz and the illumination time was $t_{\textrm{UV}}=50\,$ms. 
Here, we swept the frequency from $\pm100\,$kHz around the central frequency in order to avoid missing narrow lines between the data points.

\subsection{Correlation measurements}
For the correlation measurements, we started with a $94(1)\%$-filled Mott insulator and illuminated the cloud with UV light resonant with a macrodimer resonance until the filling decreased to $87\%$.
We took about 200 images under the same conditions.
Details on the measurements as well as the measured correlation values are shown in Table~\ref{Table:1}.
We verified that the recapture probability of our excited molecules is negligible by measuring only atoms in $F=1$ after preparing a cloud in $\ket{g}$ and shining UV light resonant with the lowest macrodimer line for various times.
Then, we perform a push-out beam resonant with the transition from the $F=2$ manifold to the D2 line.
We find that the negligible amount of $F=1$ atoms does not increase over time, while the total atom number decreases due to macrodimer excitation.
Because Rydberg atoms can decay to both hyperfine ground states, we conclude that all excited macrodimers leave the system as pairs.
We also checked that the measured correlations along the strongly coupled diagonal orientation are consistent with a reduction of the filling from the initial value to $88\%$, which is close to the observed final fillings and indicates that losses are dominated by correlated atom loss along the strongly coupled lattice diagonal. 
Finally, we have checked that the initial pair loss dynamics from the correlation measurements is consistent with the calculated Rabi coupling $\widetilde \Omega^{\pi}_0/2\pi= 3.3(6)\,$kHz to the lowest macrodimer state shown in Fig.~\ref{supfig:7} via a numerical simulation of the optical Bloch equations.


\end{document}